\documentclass[aps,preprint]{revtex4}
\usepackage{graphicx}
\begin{document}

\title{Kerr-Taub-NUT Spacetime with Maxwell and Dilaton Fields}

\author{\large Alikram  N. Aliev}
\affiliation{Feza G\"ursey Institute, P.K. 6  \c Cengelk\"oy,
34684 Istanbul, Turkey}
\author{\large Hakan Cebeci}
\affiliation{Department of Physics, Anadolu University, 26470 Eski\c{s}ehir, Turkey}
\author{\large Tekin Dereli}
\affiliation{Department of Physics,  Ko\c{c} University, 34450
Sar{\i}yer, Istanbul, Turkey }
\date{\today}

\begin{abstract}

We present an exact solution describing a  stationary and axisymmetric  object with electromagnetic and dilaton fields. The solution generalizes the usual Kerr-Taub-NUT (Newman-Unti-Tamburino) spacetime in general relativity and is obtained by boosting  this spacetime in the fifth dimension and performing a Kaluza-Klein reduction  to four dimensions. We also discuss the physical parameters of this solution and calculate its gyromagnetic ratio.

\end{abstract}


\maketitle

\newpage

\section{Introduction}

One of the remarkable generalizations of the famous  Kerr solution in general relativity is achieved  by introducing  an extra non-trivial parameter, the so called gravitomagnetic monopole moment or ``magnetic mass". The resulting solution describes the spacetime of a localized stationary and axisymmetric  object and is known as the Kerr-Taub-NUT (Newman-Unti-Tamburino)  solution of the vacuum  Einstein field equations \cite{demianski}. This solution belongs to a general class of metrics that admits separability of variables in the Hamilton-Jacobi  equations \cite{carter} and contains three physical parameters: The gravitational mass (gravitoelectric charge), the magnetic mass (NUT charge) and the rotation parameter. The presence of the NUT charge in the spacetime destroys its asymptotic structure  making it, in contrast to the Kerr spacetime, asymptotically non-flat.  Though the Kerr-Taub-NUT spacetime has no curvature singularity, there are  conical singularities  on its axis of symmetry that result in the gravitomagnetic analogue of Dirac's string quantization condition \cite{misner}. The conical singularities can be removed by imposing an appropriate periodicity condition on the time coordinate. However, the price paid for this turns out to be  the emergence of closed timelike curves (CTCs) in the spacetime that makes it hard to interpret the  solution as a regular black hole. In an alternative interpretation, one may consider the conical singularities as the source of a physical string threading the spacetime along the axis of symmetry \cite{bonnor}. In spite of these undesired features, the Kerr-Taub-NUT solution still serves as an attractive example of spacetimes with asymptotic non-flat structure for exploring various physical phenomena in general relativity \cite{aliev1, nouri, bini1,  bini2}.

The spacetimes containing a NUT charge have also played an important role in the low energy string theory where the existing duality symmetries of the effective action allowed to construct new stationary Taub-NUT type solutions \cite{jp,gk}. Moreover, due to the NUT charge a kind of new electric-magnetic duality  symmetry was revealed in supersymmetric configurations described by the Kerr-Newman-Taub-NUT-AdS solution \cite{ortin}. Some examples of gravitating solutions with NUT charge were  found in the Einstein-Yang-Mills theory \cite{radu}.

In the regime of Euclidean signature the solution with NUT charge provided one of the first examples of gravitational instantons with self-dual curvature (the Taub-NUT instanton) playing an important role in Euclidean quantum gravity \cite{h}. The NUT charge  has also led to the existence of a new type of instanton metrics with an event horizon -``bolt".  For non-rotating configurations the Taub-bolt solution was given in \cite{page} and its generalization including rotation, the Kerr-Taub-bolt instanton, was found in \cite{gperry}. For some further developments, see Refs. \cite{gibbons, hitchin, etesi, ac}.

In recent years, black hole solutions of the higher-dimensional Einstein field equations with a cosmological constant have been an  area of growing interests in the context of the AdS/CFT correspondence
\cite{hhtr,glpp}. This in turn  raised the question of their generalizations including  NUT charges. Such a generalization for a single NUT parameter and with a special restriction on the rotation parameters  was shortly found in \cite{chen1}. The most general Kerr-AdS metrics with multi-NUT charges in all higher dimensions were given in  \cite{chen2}.

In the present paper, we study the Kerr-Taub-NUT spacetime in another theoretical scheme namely, in Kaluza-Klein theory. We obtain the Kerr-Taub-NUT type solution satisfying the coupled Einstein-Maxwell-dilaton field equations which are obtained by Kaluza-Klein reduction of Einstein gravity in five dimensions. The solution turns out to be charged carrying an electric charge as well as a scalar charge (though the latter is not independent), thereby transmitting the imprints of the five-dimensional gravity into four-dimensions. The procedure of obtaining these types of solutions in the Kaluza-Klein theory relies on boosting the corresponding four-dimensional  ``seed" solutions in the fifth dimension and traces back to early works \cite{gw, fz}. (See also \cite{leu, maison, cho}).

The plan of the paper is as follows: In Sec.II we consider the field equations of the Kaluza-Klein theory  and  construct  the solution representing a boosted Kerr-Taub-NUT spacetime with associated electromagnetic and scalar potentials. In Sec.III we discuss the physical characteristics  of this solution calculating its mass, angular momenta and charges. We also discuss the electromagnetic properties of the solution  and show that it has the same gyromagnetic ratio as that for rotating Kaluza-Klein black holes obtained by boosting the Kerr metric in the fifth dimension.

\section{The solution}

The construction of the  Kerr-Taub-NUT solution with  Maxwell and dilaton fields on Kaluza-Klein reduction to four dimensions amounts to the following steps: i) one  starts with the original Kerr-Taub-NUT  metric in four dimensions and uplifts it to five dimensions by adding on an extra spacelike flat dimension, ii) next, one performs a boost in the extra fifth dimension. Clearly, the resulting metric satisfies the five-dimensional vacuum Einstein field equations, iii) finally, one performs a Kaluza-Klein reduction along the fifth dimension to four dimensions using the standard parametrization for
the five-dimensional metric
\begin{eqnarray}
ds_{5}^2 &=&   e^{-2 \Phi/\sqrt{3}}\,  ds_{4}^2 + e^{4 \Phi/\sqrt{3}}\,\left(dy + 2 A\right)^2
\label{kkmetric}
\end{eqnarray}
where the potential one-form $ A= A_{\mu}\, dx^{\mu} \,$ and $ \Phi $ is a dilaton field, the coordinates  $ x^{\mu}=\{x^0,  x^1, x^2, x^3\} =\{t, r, \theta, \phi\} $.

\subsection{Field Equations and the Seed Metric}

The simplest action that describes general relativity coupled to Maxwell-dilaton fields is given by
\begin{eqnarray}
S &=&  \int d^4 x \sqrt{-g} \left[R - 2 \left(\partial\Phi\right)^2 - e^{2 \alpha \Phi} F ^2\right]\,\,,
\label{4d action}
\end{eqnarray}
where  $ F=dA $,  and $ \alpha $ is a coupling constant. For $ \alpha=0 $, we have the usual Einstein gravity with Maxwell and scalar fields in four dimensions. The action with  $ \alpha=1 $ is a truncated action in low-energy limit of string theory.  For $ \alpha=\sqrt{3}$, we recognize in (\ref{4d action}) the action of the Kaluza-Klein theory. It is obtained from the  five-dimensional Einstein-Hilbert action through dimensional reduction procedure using an appropriate metric parametrization. In this paper we are interested only in the action  with $ \alpha=\sqrt{3} $ that gives rise to the following field equations
\begin{eqnarray}
\label{e1}
R_{\mu \nu} &=&  2 \left(\partial_{\mu} \Phi\, \partial_{\nu}\Phi + e^{2\sqrt{3} \,\Phi} \,T_{\mu \nu} \right)\,\,, \\[3mm]  &&
\nabla_{\mu}\left(e^{2\sqrt{3}\, \Phi} F^{\mu \nu} \right)= 0\,\,,
 \label{e2} \\[3mm]  &&
\nabla^2 \Phi- \frac{\sqrt{3}}{2}\,\, e^{2 \sqrt{3}\, \Phi} F^2 = 0\,\,,
\label{e3}
\end{eqnarray}
where $ \nabla_{\mu} $ stands for a covariant derivative operator and
the energy-momentum tensor of the electromagnetic field is given by
\begin{eqnarray}
T_{\mu \nu} &=&   F_{\mu \lambda}F_{\nu}^{~\lambda} - \frac{1}{4} \,g_{\mu \nu} \, F^2\,\,.
\label{emt}
\end{eqnarray}
The exact solutions for rotating black holes subject to these equations have been discussed in a number of papers. The  authors of work \cite{gw} were first to discuss the gyromagnetic ratio of rotating Kaluza-Klein black holes obtained by boosting the Kerr metric in the fifth dimension. The explicit metric for these black holes was given in \cite{fz}. Further developments can be found  in subsequent works \cite{horne,rasheed}. Here we apply the similar approach to the Kerr-Taub-NUT spacetime \cite{demianski} which in the Boyer-Lindquist coordinates is given by the metric
\begin{eqnarray}
ds_{KTN}^2 & = & -{{\Delta}\over {\Sigma}} \left(dt - \chi\,
d\phi \right)^2 + \Sigma \left(\frac{dr^2}{\Delta} +
d\theta^{\,2}\right)
+ \,\frac{\sin^2\theta}{\Sigma} \left[a  dt -
\left(r^2+a^2+\ell^2\right) d\phi\, \right]^2 , \label{4kntnut}
\end{eqnarray}
where
\begin{eqnarray}
\Delta &= & r^2 - 2M r + a^2 - \ell^2 \,\,, ~~~~~
\Sigma =  r^2 + (\ell + a \cos\theta)^2 \,\,,\nonumber \\[2mm]
\chi & = & a \sin^2\theta - 2 \ell \cos\theta \,,
\label{metfunct}
\end{eqnarray}
the parameter $ M $  is the mass of the source, $ a $ and  $ \ell $ are the rotation and NUT charge parameters, respectively. As mentioned above, the spacetime (\ref{4kntnut}) has no curvature singularities however, the presence of the NUT charge creates conical singularities on its axis of symmetry leading to Dirac-Misner string singularities at poles $\theta=0,\, \pi $.

Using the spacetime (\ref{4kntnut}) as a seed one, it is easy to go to the five-dimensional spacetime by adding on a flat fifth coordinate $ y $. This gives the metric
\begin{eqnarray}
ds_{5}^2 &=& ds_{KNT}^2 + dy^2\,,
\label{5dextension}
\end{eqnarray}
which is translationally invariant  in the direction of the fifth coordinate.

\subsection{Boosting}

Applying to the metric in (\ref{5dextension}) the boost transformation
\begin{eqnarray}
t & \rightarrow &  t \cosh\alpha + y \sinh\alpha \nonumber \\[2mm]
y & \rightarrow &  y \cosh\alpha + t \sinh\alpha
\label{boost}
\end{eqnarray}
with velocity  $ v= \tanh\alpha $, we obtain the metric
\begin{eqnarray}
ds_{5}^2 & = & -\left(1- Z \cosh^2\alpha \right)dt^2 - 2 \left(Z \chi+ 2 \ell\cos\theta \right) \left( \cosh\alpha \,dt + \sinh\alpha \,dy \right) d\phi + Z \sinh2\alpha \,dt dy \nonumber \\[3mm] &&
+ \left[\left(r^2+a^2+\ell^2\right)\sin^2\theta + \chi \left(Z \chi + 2 \ell \cos\theta \right)\right]d\phi^2  + \left(1+ Z \sinh^2\alpha \right)dy^2 \nonumber \\[3mm] &&
+ \Sigma \left(\frac{dr^2}{\Delta}+ d\theta^{\,2}\right)\,,
\label{5dboosted}
\end{eqnarray}
where we have introduced the notation
\begin{eqnarray}
Z &= &  2 \,\frac{ M r + \ell \left(\ell+ a \cos\theta \right)}{\Sigma}\,\,.
\label{zet}
\end{eqnarray}
This metric satisfies the five-dimensional Einstein field equations in vacuum, $R_{AB}=0 $, and on Kaluza-Klein reduction to four dimensions, according to the parametrization (\ref{kkmetric}), it gives the solution to the field equations in (\ref{e1})-(\ref{e3}). Thus, we have the four-dimensional solution
\begin{eqnarray}
ds_{4}^2 & = & - \frac{1-Z}{B} \,dt^2 - \frac{2 \cosh\alpha}{B}\,\left(Z \chi+ 2 \ell\cos\theta \right) dt d\phi + B \Sigma \left(\frac{dr^2}{\Delta}+ d\theta^{\,2}\right) \nonumber \\[3mm] &&
+ \left[B \left(r^2+a^2+\ell^2\right)\sin^2\theta + \frac{Z \chi + 2 \ell \cos\theta}{B} \left(\chi- 2 \ell \cos\theta \sinh^2\alpha\right)\right]d\phi^2\,\,
\label{solution}
\end{eqnarray}
with the potential one-form
\begin{eqnarray}
A &=& \frac{Z \sinh\alpha}{2 B^2}\,\left[\cosh\alpha \, dt - \left( \chi + \frac{ 2 \ell \cos\theta }{Z}\right)d\phi\right]
\label{potform1}
\end{eqnarray}
and with the dilaton field
\begin{eqnarray}
\Phi&=& \frac{\sqrt{3}}{2} \ln B\,,
\label{scalar}
\end{eqnarray}
where
\begin{eqnarray}
B &=& \left(1+ Z \sinh^2\alpha\right)^{1/2}.
\end{eqnarray}
It is useful to note that the metric (\ref{solution})  can also be written in the alternative form
\begin{eqnarray}
ds_{4}^2 & = & -\frac{1}{B}\,{{\Delta}\over {\Sigma}} \left(\,dt - \chi \cosh\alpha \,d\phi\,\right)^2 + B \Sigma \left(\frac{dr^2}{\Delta}\,+\,
d\theta^{\,2}\right) -\frac{\Delta \sin^2\theta}{B} \sinh^2\alpha\, d\phi^2
\nonumber\\[2mm] &&
+ \,\frac{\sin^2\theta}{B \Sigma} \left[a dt -
\left(r^2+a^2+\ell^2\right)\cosh\alpha\, d\phi\, \right]^2 .
 \label{solution2}
\end{eqnarray}
This metric (or that given in (\ref{solution})) represents a boosted Kerr-Taub-NUT spacetime with electromagnetic and dilaton fields. Clearly, the spacetime must be characterized not only by the mass, rotation parameter and NUT charge but also by  electric and dilaton charges. In the following section we discuss the physical parameters of the spacetime and show that the dilaton charge is indeed not an independent parameter. Taking  $\ell=0 $ in (\ref{solution}), we have the solution given in \cite{fz}.

\section{Physical Parameters}

It is straightforward to see that the asymptotic properties of the spacetime (\ref{solution}) are similar to those of  the original Kerr-Taub-NUT solution in general relativity: The NUT charge appears to be a measure of deviation from  asymptotic flatness at large distances  that is expressed by the presence of  Dirac-Misner type singularities in the off-diagonal component of the spacetime metric. However, the remaining components of the metric  turn out to have  the desired asymptotic behavior. For example, as  $ r \rightarrow\infty $, we find that
\begin{eqnarray}
g_{00} &=& -1 + \frac{M (1+ \cosh^2\alpha)}{r} + \mathcal{O}\left(\frac{1}{r^{2}}\right)\,.
\label{expg00}
\end{eqnarray}
It follows that the physical mass of the spacetime is given by
\begin{eqnarray}
\mathcal{M} &=& \frac{M}{2} \left(1+ \cosh^2\alpha \right)\,.
\label{mass}
\end{eqnarray}
We note that one can also calculate the mass using the Komar integral
over a $2$-sphere at spatial infinity
\begin{eqnarray}
\mathcal{M}  =  -\frac{1}{8 \pi\,}\oint \,^{\star}d\hat
\xi_{(t)} \,,
\label{kintegral1}
\end{eqnarray}
where $\star $ denotes the Hodge dual and the Killing one-form $ \hat
\xi_{(t)} $  is associated to the timelike isometries of the spacetime which are described  by the Killing vector $ \,\xi_{(t)}= \partial/\partial t\,$.  The direct calculations confirm the result in (\ref{mass}).

In a recent work \cite{aliev1}, it was shown that the NUT charge generates a ``rotational effect", so that the Kerr-Taub-NUT spacetime, in addition to the usual Kerr angular momentum, must be assigned a ``specific angular momentum" due to the NUT charge. Here we  follow the same strategy to calculate the angular momenta of the spacetime (\ref{solution}). Since the spacetime admits the spacelike Killing  vector  $ \,\xi_{(\phi)}= \partial/\partial \phi\,$ as well, we
start with the Komar integral
\begin{eqnarray}
J =  \frac{1}{16 \pi\,}\oint \,^{\star}d\hat
\xi_{(\phi)}\,.
\label{kintegral2}
\end{eqnarray}
Evaluating this integral  over a $2$-sphere at $ r \rightarrow\infty $, we  obtain
\begin{eqnarray}
J &=& a M \cosh\alpha \,.
\label{angmomentum1}
\end{eqnarray}
It is worth to note that though the integrand in (\ref{kintegral2}) involves an undesired term growing at  $ r \rightarrow\infty $, the result is finite and does not contain the effect of the NUT charge
(due to the integration over $ \theta $). Meanwhile, the effect of the NUT charge appears in the asymptotic form of the off-diagonal component of the spacetime  metric. We have
\begin{eqnarray}
 g_{03}& = &- 2\ell\cos\theta\cosh\alpha - \frac{2 M \cosh\alpha }{r} \left[a \sin^2\theta - \ell \cos\theta \left(1+\cosh^2\alpha\right)\right] + \mathcal{O}\left(\frac{1}{r^{2}}\right)\,\,.
  \label{expg03}
\end{eqnarray}
As we have mentioned above, the first term in this expression reflects the presence of the singularities  on the axis of symmetry. In order to be able to extract from this expression a sensible result, we first need a kind of ``renormalization" procedure removing the singular term. This is achieved, as was shown in \cite{aliev1},  by employing a background subtraction: {\it We take the difference between the off-diagonal component of the metric under consideration and that of its reference background given by $ M=0 $}. (See also works in Refs. \cite{aliev2, aliev3}). Thus, we have
\begin{eqnarray}
\delta g_{03}& = & - \,\frac{2 M \cosh\alpha }{r} \left[a \sin^2\theta - \ell \cos\theta \left(1+ \cosh^2\alpha \right)\right] + \mathcal{O}\left(\frac{1}{r^{2}}\right)\,. \label{renexpg03}
\end{eqnarray}
We see that this expression, apart from the usual Kerr-type angular momentum, also  contains a similar order contribution related to the NUT charge. To be more precise in this statement, it is useful to appeal to the twist of a timelike Killing vector. As is known, this quantity measures the failure of the Killing vector to be hypersurface  orthogonal which is in turn  determined by the rotational dynamics of the spacetime under consideration.  The twist one-form is given by
\begin{equation}
{\hat \omega}  =
\frac{1}{2}\,^{\star}\left({\hat\xi}_{(t)} \wedge
d\,{\hat\xi}_{(t)}\right)\,. \label{twist}
\end{equation}
Evaluating this one-form for the spacetime (\ref{solution}), it is easy to show that in the asymptotic region it is closed up to  order $ \mathcal{O}\left(1/r^3\right)$. That is,  as $ r\rightarrow \infty $ we have
\begin{equation}
{d\hat \omega}  = \mathcal{O}\left(\frac{1}{r^{4}}\right)\,.
\end{equation}
It follows that one can also introduce a scalar twist potential at order $ \mathcal{O}\left(1/r^2\right)$.  After performing the background subtraction procedure, as in the case of (\ref{renexpg03}), we obtain the scalar twist
\begin{equation}
\delta\Omega = \frac{M \cosh\alpha}{r^2}\left(a \cos\theta+\ell \cosh^2\alpha \right)
\label{sctwist}
\end{equation}
Comparison of this expression with that of in (\ref{renexpg03}) shows that in addition to the angular momentum in (\ref{angmomentum1}), the spacetime can be assigned an angular momentum
\begin{eqnarray}
j= M \ell \cosh\alpha \,,
\label{nutangmomentum}
\end{eqnarray}
which is related to the NUT charge of the source.

We turn now to the physical parameters which determine the electromagnetic properties of the spacetime (\ref{solution}). We calculate them exploring the asymptotic behavior of the potential one-form  (\ref{potform1}) along with  the associated field tensor components
\begin{eqnarray}
\label{emt}
F_{01} &= &-\frac{M - Z r}{2 B^4 \Sigma}\sinh2\alpha\,,~~~~
F_{13} = \frac{F_{01}}{\cosh\alpha}\, \left(\chi- 2 \ell\cos\theta\sinh^2\alpha\right)\,,\nonumber
\\[4mm]
F_{\,02}& = &-\frac{\ell \left(\Sigma- 2 r^2\right) + 2 M r \left(\ell + a\cos\theta\right)}{2 B^4 \Sigma^2}\,a\sin\theta\sinh 2 \alpha\,,\nonumber
\\[4mm]
F_{\,23} &= & \frac{F_{02}}{a \cosh\alpha}\,\left[r^2 + a^2 +\ell^2 + 2 \left(\ell^2+ M r \right)\sinh^2\alpha\right]\,.
\end{eqnarray}
The asymptotic form of the vector potential at spatial infinity is given by
\begin{eqnarray}
\label{expa0}
A_{0} &=& \frac{M \sinh\alpha\cosh\alpha}{r} + \mathcal{O}\left(\frac{1}{r^{2}}\right)\,\,,\\[4mm]
A_{3} &=& - \ell\cos\theta\sinh\alpha  -\frac{M \sinh\alpha}{r} \left(a \sin^2\theta - 2\ell \cos\theta \cosh^2\alpha \right)+
\mathcal{O}\left(\frac{1}{r^{2}}\right)\,\,.
\label{expa3}
\end{eqnarray}
Next, it is convenient to go over into orthogonal frame components of the electromagnetic field. For this purpose, we define an orthogonal frame by choosing for the metric (\ref{solution}) the following basis one-forms
\begin{eqnarray}
\label{basis1}
e^{0} &=&\left(\frac{\Delta}{B \Sigma}\right)^{1/2} \left[dt-\frac{B a \sin^2\theta- \left(Z \chi+ 2 \ell\cos\theta \right)\cosh\alpha}{1-Z}\,d\phi
\right] , \nonumber \\[4mm]
e^{3} &=&\frac{\sin\theta}{\left(B \Sigma\right)^{1/2}} \left[a\,dt -\frac{\Delta B - a \left(Z \chi+ 2 \ell\cos\theta \right)\cosh\alpha}{1-Z}\,d\phi
\right] , \nonumber \\[4mm]
 e^{1} &=&\left(\frac{B \Sigma}{\Delta
}\right)^{1/2}dr\,\,,~~~~~e^{2}=\left(B
\Sigma\right)^{1/2} d\theta\,.
\label{frame}
\end{eqnarray}
For vanishing boost parameter, $ \alpha \rightarrow 0 \,,$  this frame reduces the  well-known natural orthogonal frame in general relativity where only the radial components of the electromagnetic field survive \cite{aliev1}. However, it does not share the similar properties. It turns out that all components of the electromagnetic field do exist in this frame. Projecting the field tensor components in (\ref{emt}) onto the legs of the basis one-forms (\ref{frame}), we find the frame components of the electromagnetic field. Thus, for the asymptotic form of the electric field, we  have
\begin{eqnarray}
\label{ecomp1}
E_{\hat r} & = & F_{\hat{0}\hat{1}}= \frac{M \sinh\alpha\cosh\alpha}{r^2} + \mathcal{O}\left(\frac{1}{r^{3}}\right)\,,
 \\[4mm]
E_{\hat \theta} & = & F_{\hat{0} \hat{2}}=
\mathcal{O}\left(\frac{1}{r^3}\right)\,
\label{ecomp2}
\end{eqnarray}
and the magnetic field in the asymptotic region, $ r \rightarrow\infty \,$, are given by
\begin{eqnarray}
\label{mcomp1}
B_{\hat r} & = & F_{\hat{2}\hat{3}}= - \frac{\ell \sinh\alpha}{r^2}+ \frac{M \sinh\alpha}{r^3}\left[ 2 a \cos\theta + \ell\left( 3 \cosh^2\alpha-1\right)\right]
+\mathcal{O}\left(\frac{1}{r^{4}}\right)\,, \\[4mm]
B_{\hat \theta} & = & F_{\hat{3} \hat{1}}= \frac{a M \sinh\alpha}{r^3}\left(1-\cosh\alpha\right)\sin\theta +
\mathcal{O}\left(\frac{1}{r^4}\right)\,.
\label{mcomp2}
\end{eqnarray}
We see that both equations (\ref{expa0}) and (\ref{ecomp1}) suggest for the  electric charge the value given by
\begin{eqnarray}
Q &=& M \sinh\alpha \cosh\alpha \,\,.
\label{charge}
\end{eqnarray}
It should be noted that in contrast to the usual Kerr-Newman-Taub-NUT case \cite{aliev1}, for the boosted metric both the vector potential and the radial component of the magnetic field involve terms which  are related to the  string singularities created by the NUT charge (the first terms in equations (\ref{expa3}) and  (\ref{mcomp1})). However, these terms do not depend on the mass parameter $ M $ and therefore one can again perform a ``renormalization" by subtracting  from  each of these expressions  its value at $ M=0 $.  This yields physically meaningful expressions in the form
\begin{eqnarray}
\label{renexpa3}
\delta A_{3} &=& -\frac{M \sinh\alpha}{r} \left(a \sin^2\theta - 2\ell \cos\theta \cosh^2\alpha \right)+
\mathcal{O}\left(\frac{1}{r^{2}}\right)\,,\\[5mm]
\delta B_{\hat r} & = &  \frac{M \sinh\alpha}{r^3}\left[ 2 a \cos\theta + \ell\left( 3 \cosh^2\alpha-1\right)\right]
+\mathcal{O}\left(\frac{1}{r^{4}}\right)\,.
\label{renmcomp2}
\end{eqnarray}
These expressions along with that of in (\ref{mcomp2}) consistently show that the spacetime (\ref{solution}) has on the same footing two magnetic moments given by
\begin{eqnarray}
\mu  & = &  a M\sinh\alpha \,\,,~~~~~~~ m= \ell M \sinh\alpha \,\,.
\label{magmoments}
\end{eqnarray}

Thus, the boosted Kerr-Taub-NUT spacetime just as  the Kerr-Newman-Taub-NUT solution in ordinary general relativity  \cite{aliev1} possesses two independent magnetic  moments in accordance with its two angular momenta given in (\ref{angmomentum1}) and (\ref{nutangmomentum}). This enables us to define formally two gyromagnetic ratios by the relations
\begin{eqnarray}
 g_{1} & = & \frac{2 \mu \mathcal{M}}{J Q} \,\,,~~~~~ g_{2}  =  \frac{2 m  \mathcal{M}}{j\, Q}\,\,.
 \label{gyros}
\end{eqnarray}
With the physical parameters as given above, these relations give the equal values for the gyromagnetic ratios. That is, we have the gyromagnetic ratio
\begin{eqnarray}
g & =& 2-\tanh^2\alpha\,\,~~~~~ or ~~~~~ g  = 2-v^2
\label{gyrof}
\end{eqnarray}
which is exactly the same as that found in \cite{gw} for a Kerr black hole boosted in the fifth dimension. However, with NUT charge this value also covers the case  of vanishing Kerr parameter $ (a \rightarrow 0)$ .  We see that the gyromagnetic ratio  diminishes  from its general relativistic value $ g=2 $  to the value $ g=1 $ for charged matter in classical electrodynamics as the boost velocity varies from $ v=0 $ to its limiting value $ v \rightarrow 1 $.

We conclude this section by recalling that the spacetime (\ref{solution}) must possess a scalar charge as well. However, the scalar charge can not be thought of as an independent physical parameter.  Indeed, from the asymptotic form of the dilaton field in (\ref{scalar})
\begin{eqnarray}
\Phi&=& \frac{\sqrt{3}}{2}\, \frac{M \sinh^2\alpha}{r} + \mathcal{O}\left(\frac{1}{r^{2}}\right)\,\,,
\label{asympdil}
\end{eqnarray}
we read off the scalar charge
\begin{eqnarray}
\sigma&=& \frac{\sqrt{3}}{2}\,M \sinh^2\alpha\,\,.
\label{sccharge}
\end{eqnarray}
On the other hand, it is easy to verify that the scalar charge obeys the relation
\begin{equation}
\sigma^2 + \sqrt{3} \mathcal{M} \sigma
=\frac{3}{2}\,Q^2\,\,
\label{conrelation}
\end{equation}
that is, it can be expressed in terms  of other parameters, the mass and the electric charge.

\section{Conclusion}

In this paper, we have found a new exact solution to Kaluza-Klein theory which represents a rotating spacetime with NUT charge. We have
started with the usual  Kerr-Taub-NUT spacetime in general relativity taking it as a seed solution. Next, we have extended this spacetime to five dimensions by adding to it a flat fifth dimension.
After boosting the resulting  metric along the fifth dimension, we have performed a Kaluza-Klein reduction to four dimensions  and obtained the desired metric along with the associated electromagnetic and scalar potentials. We have given two alternative forms for this metric and calculated its physical parameters. It turned out that the new metric
retains all asymptotic features of the ordinary Kerr-Taub-NUT solution, generalizing it to include  the electric and scalar charges from five dimensions. (The scalar charge is not an independent parameter). We have shown that the new metric possesses two independent angular momenta. In addition to the usual Kerr-type angular momentum it also carries an angular momentum due to the NUT charge.

We have also defined an orthogonal frame for the metric (\ref{solution}) which for vanishing boost parameter, $\alpha \rightarrow 0 $, reduces to  a Carter-type orthogonal frame in general relativity.  This allowed us  to determine two independent magnetic moments for this the metric  and calculate its gyromagnetic ratio. The gyromagnetic ratio turned to be the same as for rotating boosted Kaluza-Klein black holes \cite{gw}. That is, it ranges between the values of 2 and 1 as the boost velocity grows from zero to its ultrarelativistic value ($ v \rightarrow 1 $).

\section{Acknowledgment}

One of us (A. N. Aliev) thanks the Scientific and Technological Research Council of Turkey (T{\"U}B\.{I}TAK) for partial support under the Research Project 105T437.

\end{document}